\documentclass[preprintnumbers,prl,superscriptaddress,twocolumn]{revtex4-1}

\usepackage{amsmath,amssymb,bm,bbm,amsfonts}
\usepackage{physics}
\usepackage{graphicx,graphics}
\usepackage[dvipsnames]{xcolor}
\usepackage[colorlinks=true,linkcolor=blue,citecolor=blue,urlcolor=blue]{hyperref}
\usepackage{dsfont}
\usepackage{comment}
\usepackage{hhline,multirow}
\usepackage{dcolumn}
\usepackage{url}
\usepackage[normalem]{ulem}
\usepackage{mathrsfs}
\usepackage{latexsym}
\usepackage{amscd}
\usepackage{slashed}
\usepackage{mathtools}
\usepackage{array}

\newcommand{\beq}{\begin{Eqarray}}
\newcommand{\eeq}{\end{eqnarray}}
\newcommand{\beqs}{\begin{eqnarray*}}
\newcommand{\eeqs}{\end{eqnarray*}}

\begin{document}

\title{On the resolution of the sign of gluon polarization in the proton}

\author{N.~T.~Hunt-Smith}
\affiliation{CSSM and ARC Centre of Excellence for Dark Matter Particle Physics, Department of Physics, University of Adelaide, Adelaide 5005, Australia}
\author{C.~Cocuzza}
\affiliation{Department of Physics, William and Mary, Williamsburg, Virginia 23185, USA}
\author{W.~Melnitchouk}
\affiliation{Jefferson Lab, Newport News, Virginia 23606, USA \\
        \vspace*{0.2cm}
        {\bf JAM Collaboration
        \vspace*{0.2cm} }}
\author{N.~Sato}
\affiliation{Jefferson Lab, Newport News, Virginia 23606, USA \\
        \vspace*{0.2cm}
        {\bf JAM Collaboration
        \vspace*{0.2cm} }}
\author{A.~W.~Thomas}
\affiliation{CSSM and ARC Centre of Excellence for Dark Matter Particle Physics, Department of Physics, University of Adelaide, Adelaide 5005, Australia}
\author{M.~J.~White}
\affiliation{CSSM and ARC Centre of Excellence for Dark Matter Particle Physics, Department of Physics, University of Adelaide, Adelaide 5005, Australia}

\begin{abstract}
Recently the possible existence of negative gluon helicity, $\Delta g$, has been observed to be compatible with existing empirical constraints, including from jet production in polarized proton-proton collisions at RHIC, and lattice QCD data on polarized gluon Ioffe time distributions. 
We perform a new global analysis of polarized parton distributions in the proton with new constraints from the high-$x$ region of deep-inelastic scattering (DIS).
A dramatic reduction in the quality of the fit for the negative $\Delta g$ replicas compared to those with positive $\Delta g$ suggest that the negative $\Delta g$ solution cannot simultaneously account for high-$x$ polarized DIS data along with lattice and polarized jet~data. 
\end{abstract}

\preprint{JLAB-THY-24-4003, \, ADP-24-04/T1243}
\date{\today}
\maketitle

{\it Introduction.---}\ 
How is the spin of the proton distributed amongst its quark and gluon constituents?
This question has perplexed physicists for almost four decades and provided motivation for experimental programs at facilities such as RHIC~\cite{Bazilevsky:2016itl}, Jefferson Lab~\cite{Burkert:2018nvj}, and the future Electron-Ion Collider~\cite{AbdulKhalek:2021gbh}.
Dedicated experiments worldwide have generally confirmed the surprising observation that quarks carry only about $\frac13$ of the total spin of the proton, with the rest presumably attributable to gluon helicity or to quark and gluon (or parton) orbital angular momentum~\cite{Thomas:2008ga}.
Significant progress was made with the availability of RHIC data on double spin asymmetries (DSAs) in inclusive jet and pion production in polarized proton-proton collisions~\cite{STAR:2014wox, STAR:2019yqm, STAR:2021mfd, STAR:2021mqa, PHENIX:2010aru}, which provided first hints for a positive polarized gluon parton distribution function (PDF), $\Delta g$, in the region above $x \approx 0.1$~\cite{deFlorian:2014yva}. 

Recently, Zhou {\it et al.}~\cite{Zhou:2022wzm} revisited the impact of RHIC spin data on the gluon polarization, with a careful examination of the theoretical assumptions that are commonly made in global QCD analysis.
Specifically, it was found that parton-level positivity constraints, which require the magnitude of polarized gluon PDF to be smaller than the unpolarized PDF, $g$, played a crucial role in determining the sign of $\Delta g$.
Relaxing these constraints revealed a possible second set of solutions in which $\Delta g$ is negative, while still describing the RHIC jet data as well as the positive solutions.
Subsequently, Whitehill {\it et al.}~\cite{Whitehill:2022mpq} demonstrated that the negative $\Delta g$ solutions can equally well describe the pion double spin asymmetry (DSA) data from PHENIX~\cite{PHENIX:2014axc, PHENIX:2020trf}.

In the absence of clear data-driven evidence ruling out the negative solutions for $\Delta g$, an alternative strategy invokes possible constraints imposed by off-the-light-cone matrix elements calculable in lattice QCD (LQCD)~\cite{Ji:2013dva, Hatta:2013gta, Zhao:2015kca, Ma:2017pxb, Ji:2020ena}.
Recently, the HadStruc collaboration performed simulations of matrix elements with direct sensitivity to $\Delta g$~\cite{HadStruc:2022yaw}, concluding that the negative $\Delta g$ solutions were significantly disfavored by their data.
Concurrently there have been growing efforts to combine LQCD and experimental data within a global QCD analysis framework~\cite{Lin:2017snn, Lin:2017stx, Bringewatt:2020ixn, JeffersonLabAngularMomentumJAM:2022aix, Gamberg:2022kdb, Barry:2023qqh}. These have illustrated that the combined information can yield stronger constraints on PDFs than those from either LQCD or experiment alone.

Karpie {\it et al.}~\cite{Karpie:2023nyg} performed such a combined analysis of spin-dependent PDFs, incorporating for the first time Ioffe time pseudo-distributions (pseudo-ITDs) from LQCD~\cite{HadStruc:2022yaw}.
They found that the inclusion of the LQCD data does not significantly alter the quality of the fits, and does not rule out the negative $\Delta g$ solutions at moderate values of $x$. 
Importantly, the effect of the LQCD data is to reduce the magnitude of the negative $\Delta g$ solutions at high $x$, and a corresponding sign change in the quark singlet, $\Delta\Sigma$, at $x \sim 0.4$, necessary to describe the polarized RHIC jet data.
As observed in Ref.~\cite{Karpie:2023nyg}, the inclusion of LQCD data induces a reshuffling between subprocesses contributing to jet DSAs, with more pronounced changes for the case of negative $\Delta g$.
In particular, a simultaneous description of the LQCD and jet data with the $\Delta g < 0$ solution leads to a negative $\Delta \Sigma$ at large~$x$, which becomes incompatible with DIS DSAs, preventing a simultaneous description of all the data (polarized DIS, jet DSAs and LQCD) with a negative $\Delta g$.

From another perspective, de~Florian {\it et al.}~\cite{deFlorian:2024utd} have argued that the negative $\Delta g$ solutions from Ref.~\cite{Zhou:2022wzm}, which violate PDF positivity at large $x$, can be ruled out by the requirement of positive cross sections for Higgs production at RHIC kinematics.
At leading order the asymmetry is directly given in terms of gluon PDFs, and a violation of gluon PDF positivity at $x \gtrsim 0.25$ could produce an unphysical Higgs production cross section~\cite{deFlorian:2024utd}.

In this Letter we revisit the question of whether existing experimental data, together with constraints from LQCD, can rule out negative $\Delta g$ solutions in a global QCD analysis of polarized PDFs.
We find that negative $\Delta g$ solutions from fits to RHIC jet and other world data cannot be ruled out by Higgs cross sections if constraints from LQCD are also included~\cite{Karpie:2023nyg}.
As we explain below, however, the combination of DSAs in polarized $pp$ jet production, lattice simulations, and fixed target high-$x$ data does strongly disfavor solutions with negative polarized glue.

{\it Observables.---}\ 
The QCD analysis of polarized and unpolarized PDFs and FFs is based on a large dataset of high-energy lepton and hadron scattering data, as described in recent JAM analyses~\cite{Cocuzza:2021rfn, Moffat:2021dji, Cocuzza:2022jye} and summarized in Table~\ref{table:chi2}. 
The unpolarized PDF sector is constrained by almost 6000 data points from inclusive DIS ($\approx 4000$ points), semi-inclusive DIS (SIDIS) ($\approx 1500$), Drell-Yan lepton-pair production ($\gtrsim 200$), and inclusive $W^\pm$, $Z$ and jet production in $pp$ and $p\bar p$ collisions ($\approx 350$ points)~\cite{Cocuzza:2021rfn}.
The description of SIDIS data requires a simultaneous determination of FFs, which are in addition constrained by single-inclusive $e^+ e^-$ annihilation (SIA) data into pions, kaons and unidentified hadrons~\cite{Moffat:2021dji}.
\begin{table*}[bt]
\caption{Comparison of the $\chi^2_{\rm red}$ values for $\Delta g > 0$ and $\Delta g < 0$ solutions across various datasets. $^*$The ``baseline'' and ``+LQCD'' cases have 365 data points in the polarized inclusive DIS dataset, compared with 1735 for the ``+ high-$x$ DIS'' set.}
\begin{center}
\begin{tabular}{p{8em}|>
{\centering\arraybackslash}p{5em}|>
{\centering\arraybackslash}p{5em}|>
{\centering\arraybackslash}p{5.7em}|>
{\centering\arraybackslash}p{5em}|>
{\centering\arraybackslash}p{5em}|>
{\centering\arraybackslash}p{5.7em}|>
{\raggedleft\arraybackslash}p{4em}}
    & \multicolumn{3}{c|}{~\boldmath{$\chi^2_{\rm red}({\Delta g > 0})$}~} & \multicolumn{3}{c|}{~\boldmath{$\chi^2_{\rm red}({\Delta g < 0)}$}~} & ~\boldmath{$N$~~~}~ \\
    \hline 
    ~\bf{Reaction} &
    baseline & + {\scriptsize LQCD} & + high-$x$~{\scriptsize DIS} & 
    baseline & + {\scriptsize LQCD} & + high-$x$~{\scriptsize DIS} \\
    \hline\hline
    ~\it Polarized & & & & & & & ~~\\
    ~Inclusive DIS & 0.95 & 0.96 & 1.21 & 0.98 & 1.12 & 1.25 & 1735$^*$~\\
    ~SIDIS & 0.85 & 0.84 & 1.08 & 0.84 & 0.96 & 1.11 & 231~~\\
    ~Inclusive jets & 0.84 & 0.89 & 0.90 & 0.88 & 1.10 & 1.44 & 83~~\\
    ~Inclusive $W^{\pm}/Z$ & 0.60 & 0.60 & 0.99 & 0.83 & 0.84 & 1.32 & 18~~\\ 
    ~\it Total & \bf 0.89 & \bf 0.90 & \bf 1.18 & \bf 0.92 & \bf 1.06 & \bf 1.24 & \bf 2067\,~\\
    \hline
    ~\it Unpolarized & & & & & & &~~\\
    ~Inclusive DIS & 1.17 & 1.17 & 1.17 & 1.18 & 1.18 & 1.19 & 3908~~\\
    ~SIDIS & 0.99 & 0.99 & 1.04 & 0.99 & 0.99 & 1.02 & 1490~~\\
    ~Inclusive jets & 1.28 & 1.28 & 1.30 & 1.29 & 1.29 & 1.30 & 198~~\\
    ~Drell-Yan & 1.21 & 1.21 & 1.21 & 1.24 & 1.24 & 1.24 & 205~~\\
    ~Inclusive $W^\pm/Z$ & 1.01 & 1.01 & 1.01 & 1.03 & 1.03 & 1.04 & 153~~\\
    ~\it Total & \bf 1.14 & \bf 1.14 & \bf 1.14 & \bf 1.15 & \bf 1.15 & \bf 1.15 & \bf 5954\,~\\
    \hline
    ~SIA & 0.86 & 0.86 & 0.89 & 0.90 & 0.90 & 0.92 & 564~~\\
    \hline
    ~LQCD & --- & 0.57 & 0.58 & --- & 1.18 & 3.92 & 48~~\\
    \hline\hline
    ~\it{Total} & \bf 1.08 & \bf 1.10 & \bf 1.13 & \bf 1.10 & \bf 1.12 & \bf 1.17 & \bf 8633\,~ \\
\end{tabular}
\label{table:chi2}
\end{center}
\end{table*}

In the spin-dependent sector we use the same datasets as in the JAM22 polarized PDF analysis~\cite{Cocuzza:2022jye}, but with a lower cut on the DIS final state mass, $W^2 > 4$~GeV$^2$, to access the high-$x$ region.
This allows us to include some 1370 additional data points on parallel and transverse DSAs from SLAC~\cite{ANTHONY1999339, ANTHONY200019, PhysRevLett.79.26, PhysRevD.54.6620, PhysRevD.58.112003, PhysRevLett.51.1135} (466 extra points), HERMES~\cite{ACKERSTAFF1997383, PhysRevD.75.012007} (35 points), and Jefferson Lab Hall~A~\cite{JeffersonLabHallA:2004tea, JeffersonLabHallA:2014gzr, JeffersonLabHallA:2014mam} (34 points), Hall~B/CLAS~\cite{PhysRevC.90.025212, PhysRevC.92.055201, PhysRevC.96.065208} (775 points), and Hall~C~\cite{SANE:2018pwx} (60 points), the latter which have never been analyzed before.
These data indirectly constrain $\Delta g$ through ${\cal O}(\alpha_s)$ corrections in the $Q^2$ evolution equations, and mixing with the quark singlet, $\Delta \Sigma$.
More direct constraints come from DSAs in jet production, which has terms that are linear and quadratic in $\Delta g$,
\begin{eqnarray}
A_{LL}^{\rm jet}(p_T,y) 
        &\propto& a_{gg} [\Delta g \otimes \Delta g] 
         + \sum_q a_{qg} [\Delta q \otimes \Delta g] \nonumber\\ 
        &+& \sum_{q,q'} a_{qq'} [\Delta q \otimes \Delta q'] 
        ~+~\mathcal{O}(\alpha_s),
\end{eqnarray}
where $p_T$ and $y$ are the transverse momentum and rapidity of the jet, and the symbol ``$\otimes$'' denotes a convolution integral. 

In addition to the experimental cross sections and DSAs fitted in the global analysis, we also consider the recent high-precision LQCD calculations of matrix elements of nonlocal operators that depend on pseudo-ITDs~\cite{Radyushkin:2017cyf}.
Pseudo-ITDs are Lorentz invariant amplitudes that can be matched to the PDFs in the $\overline{\rm{MS}}$ scheme when the invariant separation between the field operators, $z^2$, is sufficiently small~\cite{Balitsky:2021cwr}.
The gluon ($\Delta f = \Delta g$) and quark singlet ($\Delta f = \Delta \Sigma$) Ioffe time helicity distributions are defined as 
\begin{equation}
\mathcal{I}_{\Delta f}(\nu,\mu^2) 
= \int_0^1 \dd{x} x \sin(x \nu)\, \Delta f(x,\mu^2),
\end{equation}
where $\nu = p \cdot z$ is the Ioffe time and $\mu$ is the renormalization scale.
The gluon pseudo-ITD is a particularly useful observable, as it has direct sensitivity to $\Delta g$ at leading order in $\alpha_s$.
Following Karpie {\it et al.}~\cite{Karpie:2023nyg}, we include LQCD data generated on 1901 configurations of an ensemble with (2+1)-dynamical clover Wilson fermions with stout-link smearing and tree-level tadpole-improved gauge action with a lattice volume $32^3 \times 64$.
The pion mass for these configurations is $m_\pi = 358(3)$~MeV, with lattice spacing $a = 0.096(1)$~fm determined using the $w_0$ scale~\cite{BMW:2012hcm}. 
Uncertainties introduced by extrapolations to the physical pion mass are expected to be small compared with systematic uncertainties from other sources~\cite{HadStruc:2022yaw}.
In the global fit the LQCD data are treated on the same footing with experimental data, as in recent analyses that have included lattice constraints~\cite{Lin:2017stx, Bringewatt:2020ixn, JeffersonLabAngularMomentumJAM:2022aix, Gamberg:2022kdb, Barry:2023qqh, Karpie:2023nyg,Cocuzza:2023oam,Cocuzza:2023vqs}.

{\it Theoretical framework.---}\
For our theoretical analysis we employ the JAM global QCD analysis framework using fixed-order collinear factorization for PDFs and FFs, evolved at next-to-leading logarithmic accuracy in the zero-mass variable flavor number scheme with heavy quark mass thresholds $m_c=1.28$~GeV and $m_b=4.18$~GeV in the $\overline{\textrm{MS}}$ scheme~\cite{Workman:2022ynf}, and $\alpha_s(M_Z)=0.118$.
The unpolarized and polarized PDFs and FFs are parametrized at the input scale $\mu_0 = m_c$ as~\cite{Cocuzza:2021cbi},
\begin{equation}
    f(x,\mu_0) = N x^\alpha (1-x)^\beta (1 + \gamma\sqrt{x} + \eta x),
\label{e.template}
\end{equation}
with free fitting parameters $N$, $\alpha$, $\beta$, $\gamma$, and $\eta$ for each parton flavor, $f$.
For the pion and kaon FFs we set $\gamma = \eta = 0$, but keep all parameters for the unidentified hadron FFs. 
For polarized PDFs we take $\gamma = 0$, and parameterize the light quark PDFs $\Delta u$ and $\Delta d$ as a sum of a valence and a sea component.
For the sea quark $\Delta \bar{u}$, $\Delta \bar{d}$, $\Delta s$, and $\Delta \bar{s}$ PDFs we use two functions of the form~(\ref{e.template}), one of which is unique to each flavor while the other describes the low-$x$ region and is shared between all four distributions.
As in previous JAM analyses~\cite{Sato:2016tuz, Sato:2016wqj, Ethier:2017zbq, Sato:2019yez} we use data resampling, repeatedly fitting to data distorted by Gaussian shifts within their quoted uncertainties via $\chi^2$ minimization.
The resulting replica parameter sets approximate samples from the Bayesian posterior, from which PDF and FF confidence levels are computed.
We also employ the multi-step strategy, whereby data are added one step at a time as an efficient way of honing in on the minimum~$\chi^2$.

In the present analysis we consider several different fits, depending on the datasets included:
\begin{itemize}
    \item \textbf{baseline}: a simultaneous fit of unpolarized and polarized PDFs along with FFs from the most recent JAM22 analysis of helicity dependent PDFs~\cite{Cocuzza:2022jye};
    \item \textbf{+ LQCD}: includes in addition LQCD data, as outlined in Ref.~\cite{Karpie:2023nyg}, using the ``baseline'' as the prior;
    \item \textbf{+ high-}$\bm{x}$ \textbf{DIS}: includes an additional 1370 high-$x$ polarized DIS data points after reducing the $W^2$ cut from 10~GeV$^2$ to 4~GeV$^2$, while retaining the lattice data and using ``+ LQCD'' replicas as priors.
\end{itemize}
The ``baseline'' analysis of Ref.~\cite{Cocuzza:2022jye} was focused on the polarization of antiquarks in the small-$x$ region, for which a cut of $W^2 > 10$~GeV$^2$ on DIS data was sufficient.
The fit consisted of a large sample of about 1000 replicas, utilizing 146 parameters for the leading-twist PDFs and FFs, along with 81 other higher-twist and off-shell correction parameters, making for a total of 227 free parameters. 
Of these, 32 polarized PDF parameters were fitted in the ``+ LQCD'' case. 

The third scenario represents the most comprehensive JAM analysis of helicity dependent PDFs to date.
To adequately describe the additional data in the ``high-$x$ DIS'' case with a $W^2>4$~GeV$^2$ cut, we include target mass corrections (TMCs) on the twist-2 parts of the $g_1$ and $g_2$ structure functions using the Aivazis-Olness-Tung prescription~\cite{PhysRevD.50.3102}, and directly parametrize twist-3 and higher effects for $g_1$ and $g_2$~\cite{HuntSmith2024}.
This adds a further 20 parameters for a total of 52 free parameters in the ``+~high-$x$ DIS'' case.
The stability of the high-$x$ results was tested with respect to different TMC schemes~\cite{Wandzura:1977ce, Matsuda:1979ad, Blumlein:1998nv, Piccione:1997zh, Accardi:2008pc} and higher twist parametrizations (additive versus multiplicative), and comparison of the $\chi^2$ values and shapes of the PDFs and residual $g_1$ and $g_2$ functions found insignificant differences~\cite{HuntSmith2024}.

\begin{figure}[t]
\centering
\hspace*{-0.1cm}\includegraphics[scale = 0.2]{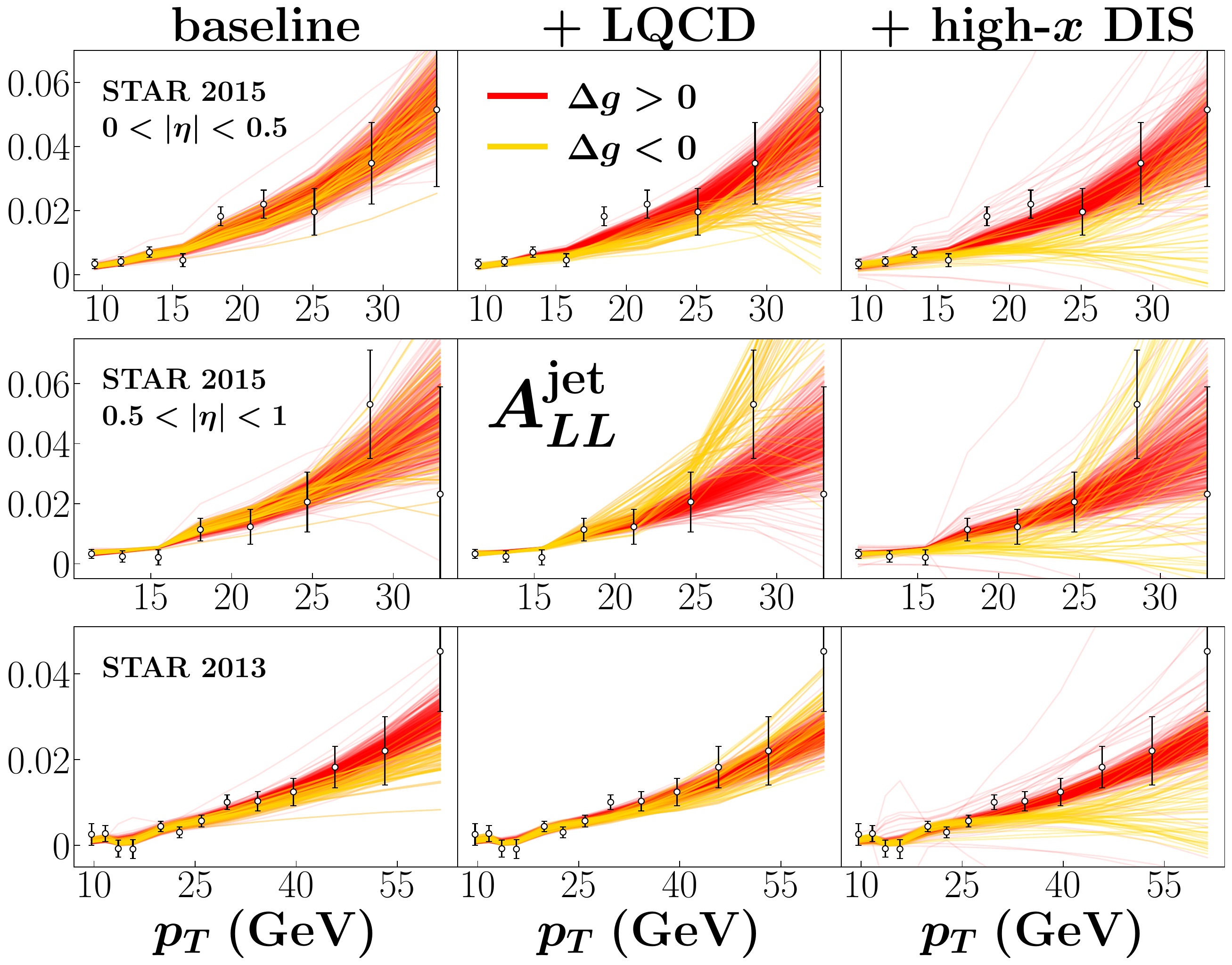}
\caption{Theory versus data comparisons for selected polarized jet DSAs $A_{LL}^{\rm jet}$ from the STAR 2015 ($\sqrt{s} = 200$~GeV)~\cite{STAR:2021mfd} and STAR 2013 ($\sqrt{s} = 510$~GeV)~~\cite{STAR:2021mqa} datasets, for positive (red) and negative (yellow) $\Delta g$ replicas. To the baseline fit (left column) are added lattice data (middle), and further additional high-$x$ polarized DIS data (right).}
\label{fig:pjet} 
\end{figure}

{\it Global QCD analysis.---}\
A comparison of the fit quality between the positive and negative $\Delta g$ replicas for each of the analyses is summarized in Table~\ref{table:chi2}, broken down into specific observables, where we give the reduced $\chi^2$ values, $\chi^2_{\rm red}$, for the different solutions.
For the ``baseline'' analysis, both the positive and negative $\Delta g$ solutions give good descriptions across all datasets, with an overall $\chi^2_{\rm red} \approx 1.1$ for each case.
This is also seen in Fig.~\ref{fig:pjet}, where a comparison of each replica set with several STAR 2015~\cite{STAR:2021mfd} and 2013~\cite{STAR:2021mqa} jet DSA datasets shows excellent descriptions for both positive and negative $\Delta g$ solutions, highlighting the limited power of the jet data to discriminate the sign of $\Delta g$~\cite{Zhou:2022wzm}.
With the addition of the LQCD data, there is a slight worsening of the $\chi^2_{\rm red}$ for both types of solutions, but no significant inconsistency for the negative $\Delta g$ replicas in either the lattice or jet DSA data, as previously observed in Ref.~\cite{Karpie:2023nyg}.

The situation changes markedly when the high-$x$ polarized DIS data are included in the fit, along with the necessary underlying theory to describe it.
While the positive $\Delta g$ replicas remain relatively stable with the addition of  each successive data set, the negative $\Delta g$ replicas yield a significantly worse fit, consistently underestimating the data at large $p_T$ values.
In particular, for the LQCD data the $\chi^2_{\rm red} \approx 0.6$ for the positive solutions remains unchanged, but increases from $\approx 1.2$ to 3.9 for the negative $\Delta g$ replicas.
A worsening is also seen in the polarized jet data, with $\chi^2_{\rm red}$ increasing  to $\approx 1.44$, and similar effects across all other polarized datasets.
We also performed an analysis including the high-$x$ polarized DIS data in the absence of the LQCD data, to check if the LQCD data were entirely necessary to see this reduction in the quality of the fit. 
In that case, there was no significant discrepancy in $\chi^2_{\rm red}$ between the positive and negative $\Delta g$ replicas for any of the datasets.
This suggests that to fully rule out the negative $\Delta g$ solution requires a simultaneous analysis of polarized jets, LQCD, {\it and} high-$x$ polarized DIS data.
\begin{figure}[tb]
\hspace*{-0.2cm}\centering\includegraphics[scale = 0.29]{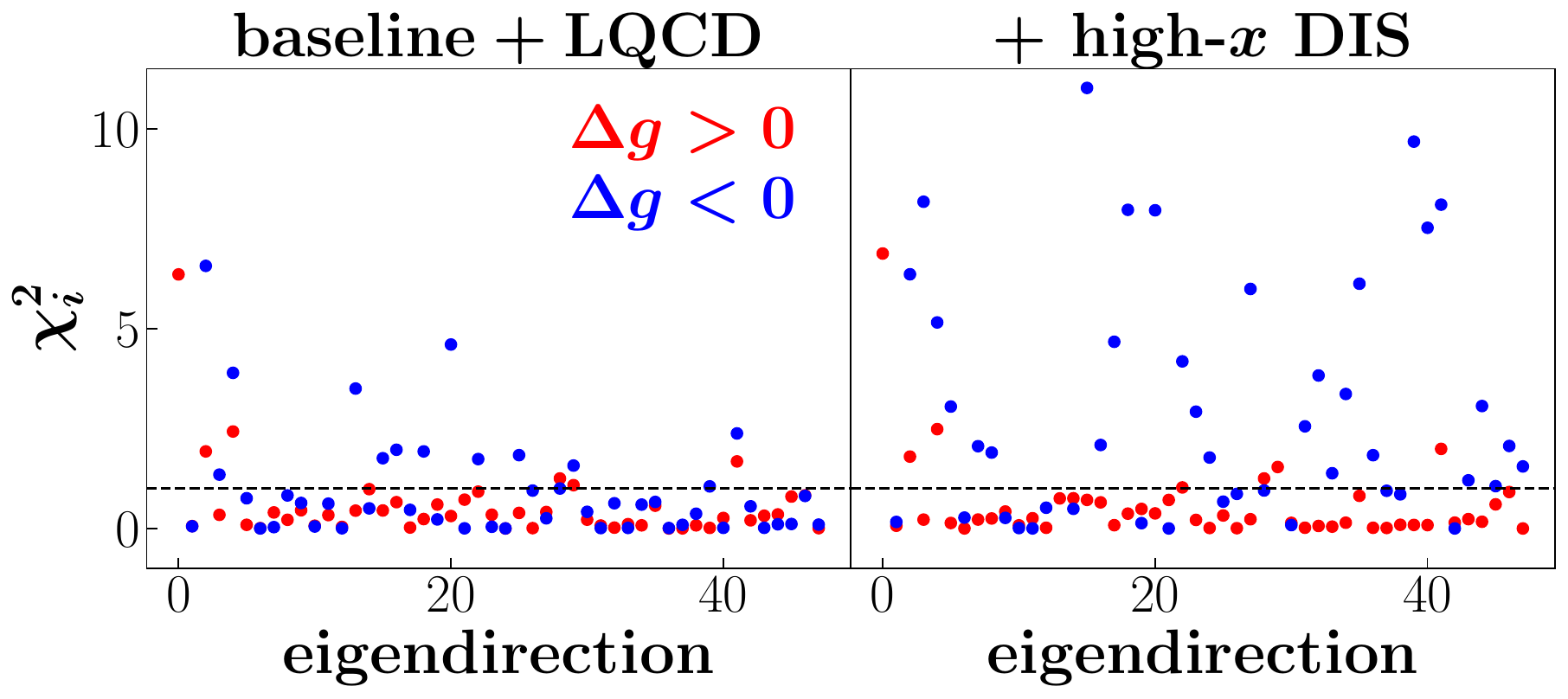}
\caption{Comparison of $\chi^2$ contributions for positive (red dots) and negative (blue dots) $\Delta g$ mean theory predictions for LQCD data versus eigendirection; $\chi^2_i$ represents the $i$-th lattice data point.}
\label{fig:lattice} 
\end{figure}

The tension between the LQCD data and the negative $\Delta g$ solution in the full fit is further illustrated in Fig.~\ref{fig:lattice}, where we show the contributions to the total $\chi^2$ from each eigendirection of the covariance matrix for the positive and negative solutions.
Since for the LQCD data we only have access to the covariance matrix between the 48 lattice points rather than individual uncertainties on each point, we rotate the residuals in the eigenspace of the covariance matrix~\cite{Karpie:2023nyg}. 
Compared with the ``baseline+LQCD'' scenario, there is clearly a larger disagreement between the theory predictions and the LQCD data across a majority of the data points after the addition of the high-$x$ DIS data.

\begin{figure}[t]
\centering
\hspace*{-0.22cm}\includegraphics[scale = 0.51]{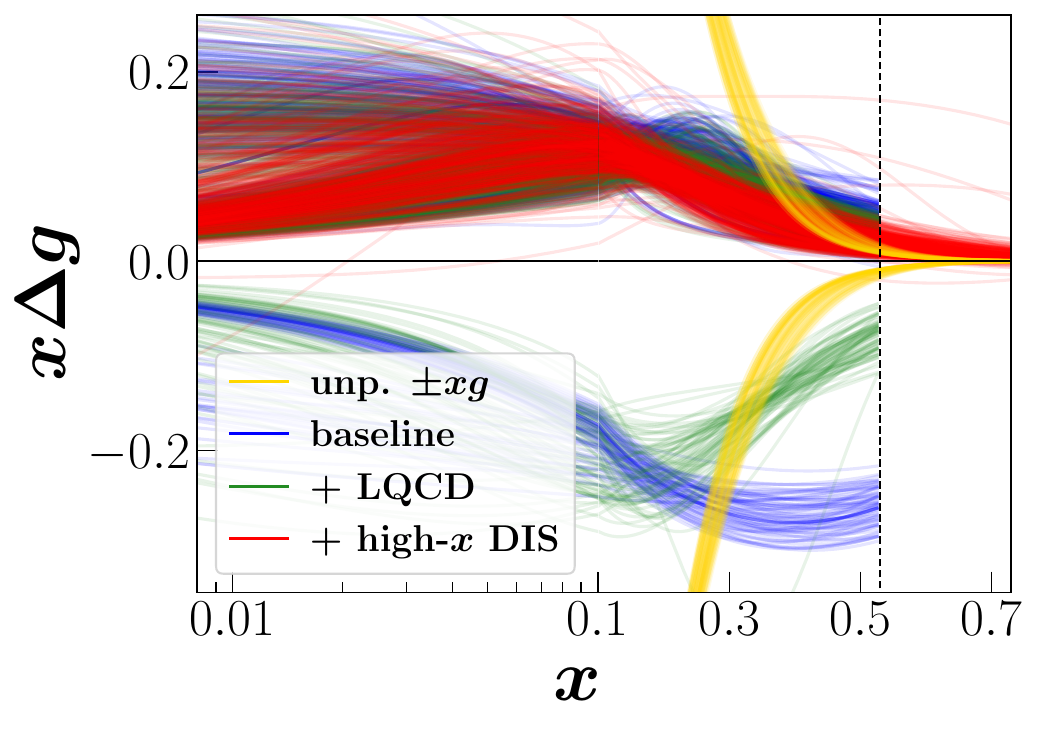}
\caption{Monte Carlo replicas for the polarized gluon PDF $x \Delta g$ and the unpolarized gluon PDF $x g$ at $Q^2 = 10$ GeV$^2$. The ``baseline'' and ``+ LQCD'' cases are cut off at $x \approx 0.53$, corresponding to a $W^2 > 10$~GeV$^2$ cut for polarized DIS.}
\label{fig:Delta_g} 
\end{figure}

Turning to the polarized gluon PDF itself, in Fig.~\ref{fig:Delta_g} we display the positive and negative $\Delta g$ solutions for each of the replica sets for the baseline, baseline + LQCD, and baseline + LQCD + high-$x$ DIS scenarios, along with the unpolarized gluon PDF for comparison. 
The negative $\Delta g$ replicas for the ``+ high-$x$ DIS'' fit are not shown, having been ruled out by the combination of all the datasets. 
By reducing the $W^2$ cut on the polarized DIS data from 10~GeV$^2$ to 4~GeV$^2$ in the ``+ high-$x$ DIS'' case, we are able to accurately characterize the polarized gluon PDF at higher values of $x$ without the need for extrapolation.
The uncertainties on $\Delta g$ are also slightly reduced following the progressive addition of the new data. 
%

{\it Higgs production.---}\
As a somewhat indirect constraint on the sign of the gluon helicity, de~Florian {\it et al.}~\cite{deFlorian:2024utd} observed that the negative $\Delta g$ replicas corresponding to the JAM22 analysis (our ``baseline'') would result in unphysical cross sections for Higgs production in polarized $pp$ collisions at RHIC, and could therefore be ruled out.
To leading order the double-helicity asymmetry for Higgs production is dominated by the gluon-gluon channel,
\begin{equation}
    \begin{aligned}
        A_{LL}^{\rm H}(\tau)\, =&
        \frac{[ \Delta g \otimes \Delta g ]}{[ g \otimes g ]}
        \, +\, \mathcal{O}(\alpha_s),
    \end{aligned}
\end{equation}
where $\tau = m_H^2/s$ for the Higgs mass $m_H$ and $\sqrt{s} = 510$~GeV is the RHIC center of mass energy.
Computing the magnitude of the asymmetry $|A_{LL}^{\rm H}|$, de~Florian {\it et al.} found this exceeded unity for the negative $\Delta g$ replicas, especially at large (unphysical) values of $m_H$.

\begin{figure}[t]
\centering
\includegraphics[scale = 0.4]{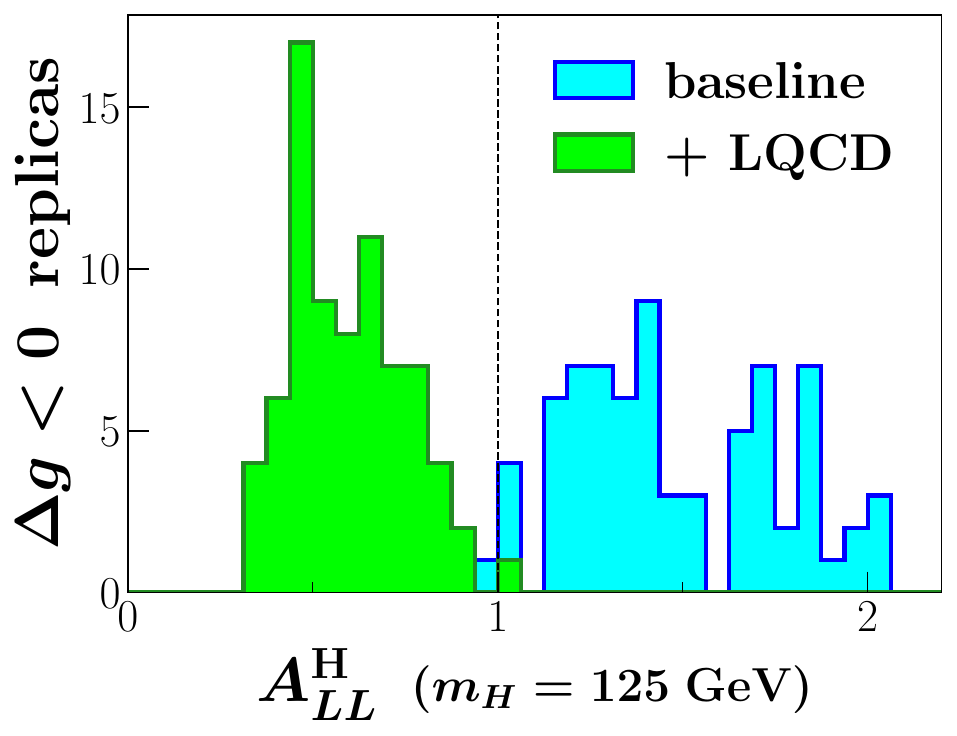}
\caption{Histogram of negative $\Delta g$ replicas for the Higgs production asymmetry $A_{LL}^{\rm{H}}$ at RHIC ($\sqrt{s} = 510$ GeV), for the ``baseline'' analysis and after the addition of LQCD data. 
}
\label{fig:A_LL_Higgs} 
\end{figure}

Repeating the calculation of $A_{LL}^{\rm H}$ at next-to-leading order as in Ref.~\cite{deFlorian:2024utd} for the physical Higgs mass, $m_H = 125$~GeV, in Fig.~\ref{fig:A_LL_Higgs}, we show a histogram of results for negative $\Delta g$ replicas for the ``baseline'' and ``+ LQCD'' scenarios.
For the ``baseline'' fit, we confirm the observations~\cite{deFlorian:2024utd} that the asymmetry exceeds unity for most of the replicas.
On the other hand, for the ``+ LQCD'' case, the vast majority of the replicas give rise to $A_{LL}^{\rm H} < 1$, so that this observable does not rule out negative $\Delta g$ values.
The negative $\Delta g$ solutions for the ``+ LQCD'' case shown in Fig.~\ref{fig:Delta_g} are therefore viable solutions which describe all of the data in the ``baseline'' fit and respect positivity constraints of the Higgs asymmetry.
It is only with the unique addition of the high-$x$ polarized DIS data described above that one is able to conclusively rule out the negative $\Delta g$ solution from the combination of empirical and lattice data, without the need for artificial constraints on the positivity of PDFs.
%

{\it Outlook.---}\
While we believe the analysis presented here reflects compelling evidence for a data-driven constraint on the sign of $\Delta g$, independent of theoretical assumptions about PDF positivity, further tests would be desirable.
For example, to complement DIS DSA data in the high-$x$ region, which are sensitive to power corrections, independent information on PDFs at large $x$ from other sources would provide better control on leading and higher twist contributions~\cite{HuntSmith2024}.
Additional lattice data on pseudo-ITDs Ioffe for the quark singlet helicity, ${\cal I}_{\Delta \Sigma}$, in addition to improved data on the gluon ${\cal I}_{\Delta g}$, would provide direct constraints.

Other observables which have been shown to have sensitivity to $\Delta g$ include DSAs in polarized SIDIS at large~$p_T$~\cite{Whitehill:2022mpq}, which could be studied at higher energies at Jefferson Lab and the Electron-Ion Collider~\cite{Zhou:2021llj}.
In addition, for dijet production in polarized $pp$ collisions at RHIC~\cite{STAR:2021mqa} it has been argued that this data can provide constraints on $\Delta g$~\cite{RHICSPIN:2023zxx}, complementing those from single jet DSAs.
The challenge there is to describe both the dijet DSAs and the unpolarized dijet cross sections~\cite{ATLAS:2013jmu} up to sufficiently large values of the dijet invariant mass, where the sensitivity of the DSAs to $\Delta g$ is strongest.
These data should provide further information on the detailed shape of the gluon helicity, and the role of gluon polarization in the proton.

\vspace*{0.3cm}

\begin{acknowledgments}

We thank J.~Karpie, C.~Monahan, K.~Orginos, J.~Qiu, D.~G.~Richards, W.~Vogelsang, and S.~Zafeiropoulos for helpful comments and discussions.
This work was supported by the DOE contract No.~DE-AC05-06OR23177, under which Jefferson Science Associates, LLC operates Jefferson Lab; and by the University of Adelaide and the Australian Research Council through the ARC Centre of Excellence for Dark Matter Particle Physics (CE200100008) and Discovery Project DP180102209 (MJW). 
The work of NS was supported by the DOE, Office of Science, Office of Nuclear Physics in the Early Career Program.

\end{acknowledgments}

\bibliography{bibliography}

\end{document}